\documentstyle[aps,prl,twocolumn]{revtex}
\begin{document}
\draft

\title{Quantum Squeezing Micron-Sized Cantilevers}
\author{M.\ P. Blencowe$^{1}$, and M.\ N.\ Wybourne$^{2}$}
\address{\it $^{(1)}$The Blackett Laboratory, Imperial College, London SW7 2BZ,
 United Kingdom}
\address{\it $^{(2)}$Department of Physics and Astronomy, Dartmouth College,
Hanover, New Hampshire 03755-3528} 
\date{\today}
\maketitle
\begin{abstract}
We show that substantial quantum squeezing of mechanical motion can 
be achieved for micron-sized cantilever devices fabricated using available
techniques. A method is also described for measuring the cantilever
fluctuation magnitudes in the squeezing regime.     
\end{abstract}

Squeezed states---minimum uncertainty states of a harmonic system where the
uncertainty of one of the quadrature amplitudes is  reduced below that of the 
zero-point fluctuations (i.e., ground state)---first came to prominence in
the late seventies and early eighties as a means to suppress
noise in optical communications\cite{yuen} and in interferometric\cite{caves1}
and mechanical bar gravity wave detectors\cite{hollen,caves2,grish}. 
The first experimental demonstration of squeezed light states   
followed shortly thereafter\cite{slusher}.
Many other groups have since  demonstrated
squeezed light using a variety of generation and detection 
techniques (see, e.g., Ref.\ \cite{kimble} for a  survey  
 up to 1992). By contrast, there has been very little experimental
work on squeezed states in {\it mechanical} systems; squeezed states have been
demonstrated for a single, vibrating ion\cite{meek} and possibly for crystal 
phonons\cite{garrett}; there have also been several theoretical 
proposals\cite{nori}. It would be of great interest to try to produce
squeezed states for a mechanical oscillator structure much larger than a single
atom, not only to test some of the ideas developed for the detection of
very weak forces such as gravity waves (see, e.g., Ref.\ \cite{braginsky}), 
but also, at a more fundamental level, to
extend the domain of manifestly quantum phenomena to macroscopic mechanical
systems (see, e.g., Ref.\ \cite{bose} for a recent proposal to generate
and detect quantum superpositions of spatially separated states in a
macroscopic mechanical system).

One way  to squeeze a mechanical oscillator
initially in a thermal state would be to use parametric pumping, characterized
by a term of the form $P(t) (a^{\dag 2} +a^2 )$ in the oscillator
 Hamiltonian (see, e.g.,
Ref.\ \cite{grish}). The first demonstration of this method for {\it classical}
thermomechanical noise squeezing was performed by 
Rugar {\it et al.}\cite{rugar} using a device 
comprising a cantilever several hundred microns in 
length and a few microns in crossection. The room temperature
 thermal vibrational  motion in the
fundamental flexural mode was parametrically squeezed in one quadrature to
an effective temperature of about 100~K by 
periodically modulating the effective spring constant at twice the flexural 
frequency.   A natural question to ask is whether {\it quantum} squeezing
could be achieved in a similar device. In order to squeeze below the zero-point
fluctuations,  the thermal fluctuations of the cantilever before squeezing
must not be too much larger in magnitude  than the zero-point 
fluctuations\cite{grish}. Now, the lowest temperature to which  a 
microdevice can be cooled using reasonably available equipment
(e.g., a nuclear demagnetization cryostat)
is around  a mK. At such    temperatures,
we would require a cantilever with a  fundamental frequency of around
100~MHz. A cantilever vibrating at radio frequencies might seem hopelessly
unrealistic. However,  recently
Cleland {\it et al.}\cite{cleland} succeeded in fabricating {\it micron}-sized,
suspended Si beams with fundamental resonant frequencies of just  this order. In 
this paper, we will show  that substantial  quantum squeezing can in fact  
be achieved  using a cantilever device similar to that of Rugar {\it et al.}\cite{rugar},
but scaled-down to micron dimensions and with materials characteristics such 
as quality factor 
similar to those of the structures considered by Cleland {\it et al.}\cite{cleland}.
We will first discuss the method of generation and then follow with the 
method of detection.

The model structure which we consider is similar to the device of Rugar 
{\it et al.}\cite{rugar},
comprising a cantilever with one  plate of a capacitor located 
on the cantilever surface and the other plate located on the substrate surface 
directly opposite. 
Unlike  their device, however, the capacitor will serve as
 a component not only of 
the pump
circuitry, but of the probe circuitry as well (more on this later). The 
classical equations of motion for the cantilever in the fundamental flexural
mode are        
\begin{equation}
\frac{d^2 x}{d t^2}+\frac{\omega_c}{Q_c}\frac{d x}{d t}+\omega_c^2 x=
\frac{q^2(t)}{2 C_0 d m} +\frac{F_{\rm fluct}(t)}{m},
\label{eqnmotion1}
\end{equation}  
where, in terms of the pump voltage $V_p(t)$, the capacitor plate charge is
\begin{equation}
  q(t)=\frac{C_0 V_p(t)}{1-x(t)/d}.
\label{charge}
\end{equation}
The coordinate $x$ denotes the displacement of the cantilever tip from
the static  equilibrium position ($V_p=0$), $m$ is the cantilever effective mass,  $d$
is the equilibrium  cantilever tip-substrate base separation,  and
$C_0$ is the  capacitance for equilibrium separation $d$. Recall that,  in
terms of the frequency $\omega_c$ and relaxation time $\tau_c$ of the 
fundamental flexural mode, the quality factor is defined  as 
$Q_c=\omega_c \tau_c$. Both $Q_c$ and the random force 
term $F_{\rm fluct}$ model the effects of the thermal environment on the 
flexural mode.  The $c$  subscript, which denotes `cantilever', is employed 
in order  to distinguish the mechanical oscillator from
the coupled probe  electrical oscillator  to be introduced later on.

Substituting (\ref{charge}) into (\ref{eqnmotion1}) for pump voltage having 
the form $V_p(t)=V_0 \cos (\omega_p t +\phi)$ and assuming $|x|\ll d$, we
obtain
\begin{equation}
m_c\frac{d^2 x}{d t^2}+\frac{m_c\omega_c}{Q_c}\frac{d x}{d t}+[k_0 +k_p (t)]
 x\approx F_p(t) +F_{\rm fluct}(t),
\label{eqnmotion2}
\end{equation}
where  $k_0=m_c\omega_c^2 +\Delta k$, $\Delta k= C_0 V_0^2/2d^2$,   
$k_p(t)=\Delta k  \cos(2\omega_p t+2\phi)$, and $F_p(t)\equiv k_p(t)$. Note that the 
equilibrium static
spring  constant is shifted upwards by $\Delta k$. Thus, the resonant frequency
of the cantilever is shifted to $\omega'_c =\sqrt{\omega_c^2 +\Delta k/m}$. 
There
is also a shift downwards in the equilibrium position of the cantilever tip
by the  amount $ C_0 V_0^2/(2dm_c \omega_c^2)$ and we have redefined the
origin of $x$  to coincide with  this new equilibrium position. Note that 
one consequence of applying the pump voltage $V_p(t)$
across the capacitor is the sinusoidal modulation $k_p(t)$   of the spring 
constant. For  phase $\phi=-\pi/4$, this modulation causes squeezing in the
quadrature amplitude $X_1$\cite{grish,rugar}, where  
\begin{eqnarray}
X_1(t) & = & x(t)\cos\omega'_c t -\omega'^{-1}_c{\dot{x}(t)}\sin\omega'_c t 
\nonumber\\
X_2(t) & = & x(t)\sin\omega'_c t +\omega'^{-1}_c{\dot{x}(t)}\cos\omega'_c t. 
\label{quadrature}
\end{eqnarray} 
Pumping the cantilever from an initial thermal equilibrium state at
frequency $\omega_p=\omega'_c$, one obtains for the quantum 
uncertainty in $X_1$\cite{grish}
\begin{equation}
\Delta X_1^2(t\rightarrow\infty)\approx\frac{\hbar}{2 m \omega_c}(2 n_T +1)
\left(1 +\frac{ Q_c\Delta k}{2 m \omega^2_c}\right)^{-1},
\label{uncertainty}
\end{equation}
where $n_T =(e^{\hbar\omega_c/k_B T}-1)^{-1}$. Note that we have 
replaced $\omega'_c$ with $\omega_c$ in (\ref{uncertainty}) since this causes
only a small error  for the parameter values to be considered below. 
On the other
hand, it is important to account for the frequency shift $\omega'_c -
\omega_c$ when setting the pump frequency $\omega_p$,  since the resonance width
$\omega'_c/Q_c$ can be smaller than this shift for large $Q_c$.
In order to have quantum squeezing, we require that the squeezing factor 
$R=\Delta X_1/\sqrt{\hbar/2 m\omega_c}<1$, where recall that 
$\sqrt{\hbar/2 m\omega_c}$ is the zero-point
uncertainty. Thus, from (\ref{uncertainty}) we have
\begin{equation}
R=\sqrt{\frac{2 n_T +1}{1 + Q_c\Delta k/2 m \omega^2_c}}<1.
\label{squeeze}
\end{equation}

For illustrative purposes, we consider a crystalline sapphire cantilever 
with mass density $\rho=3.99\times 10^3~{\rm kg/m}^{3}$ and
assume the bulk value for Young's modulus: $E=3.7\times 10^{11}~{\rm N/m}^{2}$.
Sapphire is elastically isotropic to good approximation, 
thus simplifying the analysis. The preferred  
materials for experiment would probably be  Si or GaAs. Substituting
into (\ref{squeeze}) the expressions for the cantilever effective mass,
$m=\rho l w t/4$, fundamental flexural frequency, $\omega_c =3.516 
\sqrt{E/12\rho}\ t/l^2$, and capacitance, $C_0 =\epsilon_0 \lambda l w/d$, we
obtain the following conditions on the cantilever dimensions 
\begin{equation}
\frac{t}{l^2}\lesssim\frac{1}{75}
\label{condition1}
\end{equation}
\begin{equation}
\frac{l^4}{d^3 t^3}\gtrsim 10^7,
\label{condition2}
\end{equation}    
where we have set $V_0=1$~V, $Q_c =10^4$, $T=1$~mK, and have expressed the 
various
cantilever dimensions in units of micrometers. The parameter $\lambda$, which
appears in the expression for the capacitance, is a geometrical factor to 
allow for the fact that the capacitor plate area need not 
coincide with the cantilever area. We have arbitrarily set $\lambda=1/3$.
Condition (\ref{condition1}) follows from requiring that the thermal occupation
number $n_T$ be of order one or smaller, while condition (\ref{condition2})
follows from requiring
that the term appearing in the denominator in (\ref{squeeze}) be of
order one or larger. We must  have condition (\ref{condition1}) as well since, if 
it did not hold, then $n_T$ would be very large and unrealisable
values for $Q_c$ and the various cantilever dimensions would be required in 
order to compensate. Setting $l^2/t=75$ and substituting into 
(\ref{condition2}), we obtain $d^3 t \lesssim 6 \times 10^{-4}$. 
As an example, the conditions are satisfied if we choose 
$d=0.05~\mu{\rm m}$, $t=0.1~\mu{\rm m}$, and $l=\sqrt{7.5}\approx 
2.7~\mu{\rm m}$. For these cantilever dimensions, the squeezing factor 
(\ref{squeeze}) is $R\approx 0.25$ and, thus, we have quantum squeezing.
This factor is comparable with the best
squeezing factors achieved for light\cite{kimble} and 
corresponds to a noise reduction more than three orders of
magnitude larger than that  obtained in the recent phonon squeezing 
experiments of
Garrett {\it et al.}\cite{garrett}.

If a cantilever could be realised having the same dimensions, but with
quality factor $Q_c \approx 10^6$ instead of $10^4$, then the squeezing factor
would be $R\approx 0.025$, an order magnitude smaller. It clearly would be
of great interest to determine whether micron-sized cantilevers could be
fabricated with much 
higher quality factors. Little is currently known about the   upper
limits on $Q_c$. The few reported $Q_c$ values for  micron-sized 
cantilevers\cite{cleland,stowe} are many orders of magnitude smaller than what 
can be achieved for large-scale mechanical resonators\cite{braginsky2}. 
This is thought to be due to the increasing importance of surface defects for
the dissipation of mechanical energy the larger the surface-to-volume ratio.
Presumably, the surface defect density can be reduced with appropriate 
modifications of the fabrication process.

In the analysis above, it was assumed that the cantilever tip displacement $x$
is much smaller in magnitude than the cantilever-substrate separation $d$.
One might worry, however, that this is not the case for the very large 
electric fields resulting from applying 
a potential difference of 1~V across a gap $d=0.05~\mu{\rm m}$. 
Substituting the
various parameter values into the expression for the  equilibrium 
position shift, we obtain a displacement of about $3$~\AA\ which is much smaller
than the cantilever length. We are therefore far from snapping the cantilever
and well within the range of applicability of  Hooke's force law. 
Similarly, the applied force $F_p (t)$ gives a displacement
amplitude of about $1$~\AA\ for $\omega_p=\omega'_c$. Note that, if the
frequency of the applied force was resonant with the frequency $\omega'_c$
(instead of being twice this frequency), then the displacement amplitude
would increase by a factor $Q_c =10^4$ to about $1~\mu{\rm m}$.  
We can now see that 
the displacement amplitude is small, despite the large applied electric field, 
because the applied force is off-resonance.
It is also important
to check that the applied force $F_p (t)$ is not resonant with a higher flexural
mode of the cantilever. The frequency of the second flexural mode  is about 
six times
larger than the fundamental frequency and therefore the applied force is even
further off-resonance with this  mode. An estimate of the resulting 
displacement amplitude yields about $0.1$~\AA.

The Casimir force can also give rise to large deflections for submicron plate
separations\cite{serry}. Using the expression for the Casimir force between
two parallel plates of area $A$, $F_{\rm casimir}=\pi^2\hbar c A/240 d^4$, 
we obtain a
deflection on the order of an Angstrom for a cantilever with the above dimensions, 
including a width $w=1~\mu$m.

In the classical squeezing analysis of Rugar {\it et al.}\cite{rugar}, the
analogous quantity to the squeezing factor (\ref{squeeze}) is the gain, 
defined as $G=|X|_{\rm pump\ on}/|X|_{\rm pump\ off}$, 
where $|X|=\sqrt{X_1^2 +X_2^2}$. The term $Q_c\Delta k/2 m \omega^2_c$
[see (\ref{squeeze})] also appears in their expression for $G$. However,
there would appear to be a discrepancy:  their solutions for 
$X_{1\atop 2}(t)$ break down when this term exceeds one, hence restricting their
squeezing maximum (minimum gain) to 1/2, whereas we have  no upper bound
on this term. The resolution lies in the fact that
Rugar {\it et al.} assumed steady-state solutions. If this term exceeds
one, as is the case for the parameter values we are considering, then $X_2 (t)$
grows exponentially without bound as $t\rightarrow\infty$. Thus, the pumping
should terminate after the characteristic time $t_{\rm ch}$  
for which the squeezing factor
largely reaches its limiting value (\ref{squeeze}), where\cite{grish}
\begin{equation}
t_{\rm ch}/\tau_c=\frac{2m\omega_c^2}{Q_c \Delta k}\ln
\left(\frac{Q_c \Delta k}{2m\omega_c^2}\right).
\label{characteristic}
\end{equation}

A related  issue concerns the conversion of mechanical energy into heat, 
possibly warming the cantilever sufficiently to take
it out of the quantum squeezing regime. It is not clear whether the generated 
heat would dissipate sufficiently rapidly into the surrounding substrate to
prevent this from happening; the heat dissipation rate clearly depends on the 
materials properties
and layout of the device. Alternatively, heating will be negligible if the
cantilever is pumped for a time smaller than the relaxation time. Therefore,
we require $t_{\rm ch} <\tau_c$. Substituting into (\ref{characteristic}) 
the various chosen parameter values, we find $t_{\rm ch}/\tau_c\approx 0.1$.
Thus, the limiting squeezing value can be largely attained without significant
heating.

We now consider a possible way to measure the uncertainty $\Delta X_1$. For the
considered parameter values, the zero-point uncertainty is 
$\sqrt{\hbar/2 m\omega_c}\approx 4\times 10^{-14}$~m, while the squeezed 
uncertainty is $\approx 10^{-14}$~m. Measuring fluctuations of this small
magnitude might at first appear a hopeless task. However, compare these
numbers to the even smaller displacements of around $10^{-19}$~m which
must be resolved for a metre long gravity wave bar detector. 
Remarkably, a displacement 
detector with sensitivity approaching $10^{-19}$~m was demonstrated as long
ago as 1981\cite{braginsky3} (see also Sec.\ 10 of Ref.\ \cite{braginsky2}).
The detector was effectively an  $LC$ circuit, where  changes
in the capacitor plate separation are converted into changes in the resonant
frequency $1/\sqrt{LC}$ of the circuit. This method is well-suited to our
cantilever system, since the  existing capacitor can also 
be used as a displacement sensor by forming part of an $LC$
circuit. As an aside, note that capacitance-changes have also been used to
detect cantilever deflection in an AFM (see, e.g., Ref.\ \cite{neubauer}).

Thus, we have in mind a two-stage process. In the pumping stage, 
 the cantilever flexural mode vibrations are 
driven into a squeezed state as described above. After a time $t_{\rm ch}$,
pumping terminates and then the probe circuitry takes over.     
Forming an $LC$ circuit with the existing capacitor, the cantilever equation
of motion (\ref{eqnmotion1}) becomes one of two coupled equations, where
the other equation for the circuit charge $q$ is
\begin{equation}
\frac{d^2 q}{d t^2}+\frac{\omega_e}{Q_e}\frac{d q}{d t}+\omega_e^2 
q\left(1-\frac{x}{d}\right)=
\omega_e^2 C_0 \left[V_{\rm pr}(t) +V_{\rm fluct}(t)\right].
\label{eqnmotion3}
\end{equation}        
The circuit resonant frequency is $\omega_e =1/\sqrt{LC_0}$, where 
the $e$ subscript stands for `electrical'. 
We have also included a fluctuating voltage $V_{\rm fluct}(t)$, which 
describes the Johnson-Nyquist noise due to  unavoidable circuit
resistance $R=1/(\omega_e Q_e C_0 )$.

For a continuous measurement, it is
essential that $X_1 (t)$ is measured and not the ordinary displacement $x(t)$. 
In the latter case, the Heisenberg uncertainty principle would prevent one from
measuring displacements below the zero-point fluctuations\cite{braginsky}. 
The quadrature $X_1 (t)$ is measured for   probe voltage satisfying   
$V_{\rm pr}(t)=V_0\cos\omega_e t\cos\omega_c t$. This form 
couples $q$ much more strongly to the quadrature amplitude $X_1$ than to
$X_2$, provided  $Q_e\omega_c/\omega_e\gg 1$ (see Sec.\ 10.7 of Ref.\ 
\cite{braginsky}). The necessary large  quality factors $Q_e$ can be  achieved, 
for example, by using superconducting wires 
(see, e.g., Sec.\ 6 of Ref.\ \cite{braginsky2}).

Simplifying (\ref{eqnmotion1}) and (\ref{eqnmotion3}) 
as in Sec.\ 10.5 of Ref.\ \cite{braginsky}
to make them approximately linear in  $q$ and $x$, and then solving the
corresponding quantum equations motion along the lines of Ref.\
\cite{grish},  the uncertainty in the voltage across the 
capacitor for times $1/\omega_e\ll t\ll{\rm min}(\tau_c,\tau_e)$ is 
\begin{eqnarray}
\Delta &V&^2(t)\approx\frac{1}{8}\left(\frac{\Delta X_1 (0)}{d}\right)^2 
\left(V_0\omega_e t\right)^2
+\frac{\hbar\omega_e}{2 C_0} \left(2 n_T^e +1\right)\nonumber\\
&&+\frac{1}{24}\left(\frac{\hbar}{2 m \omega_c d^2}\right)
\left(V_0\omega_e t\right)^2 
\left(\frac{\omega_c t}{ Q_c}\right)   
\left(2 n_T^c +1\right),
\label{voltfluct}
\end{eqnarray}
where $\Delta X_1(0)$ is the cantilever quadrature uncertainty at the start
of the probe stage, defined as $t=0$. The circuit is assumed to be
initially in thermal equilibrium, described by the distribution
$n_T^e=(e^{\hbar\omega_e/k_B T}-1)^{-1}$.  The right hand side of 
(\ref{voltfluct}) involves three terms,  the first of which depends
directly on $\Delta X_1(0)$. This first term increases with time, eventually
exceeding the second term which describes the voltage  fluctuations across the
capacitor due to the nonzero circuit resistance. On the other hand, 
the third term, which describes the return to thermal equilibrium of the
cantilever because it is  no longer being pumped, eventually exceeds
the first term.   
Thus, the time interval over which the uncertainty $\Delta X_1 (0)$ can be 
resolved is bounded both above and below by thermal noise.  
Choosing $\omega_e=\omega_c$, $Q_e >Q_c$, $V_0=1$~V, and the above considered
cantilever parameter values, we find $10<\omega_e t<850$, with 
$\Delta V(t)\approx\Delta X_1 (0) V_0 
\omega_e t/2\sqrt{2}d\approx 6\times 10^{-5}$~V at, say, $\omega_e t=800$
for $\Delta X_1 (0)=10^{-14}$~m. 
Again, if a cantilever with larger quality factor $Q_c\approx 10^6$ 
 could be realised (and also  $Q_e >Q_c$), 
 then the upper bound would increase to $8.5 \times 10^4$, giving a
 much larger signal $\Delta V(t)\approx 6\times 10^{-3}$~V at 
$\omega_e t=8\times 10^4$.

The uncertainty $\Delta X_1 (0)$ is a quantum statistical quantity:
the pump and probe stages must be repeated many times in order to accurately
determine                   
$\Delta X_1 (0)$. Thus, we require not only frequency stability 
but also amplitude
stability for the applied voltages $V_p (t)$ and $V_{\rm pr}(t)$. 
In particular, during the pump stage the shift in the 
equilibrium position of the cantilever tip due to uncontrollable fluctuations 
in $V_p(t)$ must be smaller than the squeezed
magnitude $10^{-14}$~m obtained above. For the considered values,
this imposes the requirement $|\delta V_p/V_p|<10^{-5}$. 

An implicit assumption of  our analysis is that the capacitor plate on
the substrate surface 
opposite the cantilever is rigidly fixed, with negligible fluctuations 
as compared
with those of the cantilever tip---even in the squeezing regime. This must be
checked, of course. We have obtained estimates of the surface fluctuations,
modelling the capacitor plate/substrate surface as a half space\cite{ezawa}. 
Summing over all the different  mode contributions 
(e.g., Rayleigh, mixed P-SV, etc.)  
to the surface fluctuations averaged over a plate area $1~\mu{\rm m}^2$, 
we find a fluctuation magnitude perpendicular to the surface of about 
$3\times 10^{-15}$~m at 1~mK. This figure is indeed smaller
 than the squeezed magnitude value $10^{-14}$~m. 
 Note, however,  that these
surface fluctuations will prevent one from measuring squeezing amplitudes 
not much below this value for the current setup.

In conclusion, we have shown that substantial squeezing  below the zero-point
motion can be achieved for the flexural mode of a micron-sized cantilever.
A method using a coupled $LC$ circuit to measure the 
fluctuation amplitudes in the squeezing regime   was also described.
Displacement detectors  with much better 
sensitivity than required were demonstrated some
time ago\cite{braginsky2,braginsky3}, while 
cantilevers with the required
dimensions and materials characteristics have recently been 
realised\cite{cleland}.

We thank S.\ Bose and D.\ A.\ Williams for helpful and stimulating discussions.

\end{document}